\documentclass[prl,aps,nobalancelastpage,showpacs,amssymb,citeautoscript]{revtex4-1}

\usepackage{graphicx,multirow}
\makeatother

\begin{document}

\title{Superconductivity, Electronic Phase Diagram and Pressure Effect in Sr$_{1-x}$Pr$_{x}$FBiS$_{2}$}

\author{Wei You$^{1}$, Lin Li$^{1}$, Haiyang Yang$^{1}$, Jialu Wang$^{1}$, Hongying Mao$^{1}$, Li Zhang$^{5}$, Chuanying Xi$^{3}$, Jie Cheng$^{4}$, Yongkang Luo$^{2}$, Jianhui Dai$^{1,*}$\footnote[1]{Electronic address: daijh@hznu.edu.cn}, Yuke Li$^{1,*}$\footnote[1]{Electronic address: yklee@hznu.edu.cn}
\\}

\address{$^{1}$ Department of Physics and Hangzhou Key Laboratory of Quantum Matter, Hangzhou Normal University, Hangzhou 310036, China}
\address{$^{2}$ Wuhan National High Magnetic Field Center School of Physics, Huazhong University of Science and Technology, Wuhan 430074, China}
\address{$^{3}$ High Magnetic Field Laboratory, Chinese Academy of Sciences, Hefei 230031, China}
\address{$^{4}$ College of Science, Center of Advanced Functional Ceramics, Nanjing University of Posts and Telecommunications, Nanjing,
Jiangsu 210023, China}
\address{$^{5}$ Department of Physics, China Jiliang University, Hangzhou 310018, China}

\date{\today}

\begin{abstract}

Based on a combination of X-ray diffraction, electrical transports, magnetic susceptibility, specific heat and pressure effect measurements, we report a series of BiS$_{2}$-based Sr$_{1-x}$Pr$_{x}$FBiS$_{2}$ superconductors with the maximum $T_c$ of 2.7 K for $x =$ 0.5 at ambient pressure. Superconductivity can be only observed at 0.4 $\leq$ x $\leq$ 0.7 while the normal state resistivity shows the semiconducting-like behaviors. $\chi(T)$ displays the low superconducting shielding volume fractions and $C(T)$ does not exhibit distinguishable anomaly around $T_c$, suggesting a filamentary superconductivity in the Pr-doped polycrystalline samples. By varying doping concentrations, an electrnoic phase diagram can thus be established. Upon applying pressure on the optimal doping Sr$_{0.5}$Pr$_{0.5}$FBiS$_{2}$ system, $T_c$ is abruptly enhanced, reaches 8.5 K at a critical pressure of $P_c =$ 1.5 GPa and remains a slight increase up to 9.7 K up to 2.5 GPa. Accompanied with the enhancement of superconductivity from the low-$T_c$ phase to the high-$T_c$ phase, the normal state undergoes a semiconductor-to-metal transition under pressure. The scenario can be linked to the enhancement of orbital overlap of Bi-6p and S-p, contributing to the bulk superconductivity above $P_c$. Phase diagram for Sr$_{0.5}$Pr$_{0.5}$FBiS$_{2}$ is also obtained.

\end{abstract}

\maketitle

\section{\label{sec:level1}Introduction}

The discovery of superconductivity with \emph{T$_{c}$} of 8.6 K\cite{BOS} in a layered crystal structure compound Bi$_{4}$O$_{4}$S$_{3}$ has evoked considerable attention. Following this work, many BiS$_{2}$-based superconductors including \emph{Ln}O$_{1-x}$F$_{x}$BiS$_{2}$(\emph{Ln}=La, Ce, Pr, Nd)\cite{LaFS,NdFS,LaFS2,CeFS,PrFS} with the highest \emph{T$_{c}$} of $\sim$10 K have been then reported and studied. Similar to the copper oxide and the iron-based superconductors with alternating stacks of superconducting and blocking layers, the crystal structure of BiS$_2$-based compounds is composed of the common superconducting BiS$_2$ layers intercalated by various block layers, e.g., Bi$_{4}$O$_{4}$(SO$_{4}$)$_{1-x}$ or [Ln$_{2}$O$_{2}$]$^{2-}$. Subsequently, another family of BiS$_2$-based superconductors Sr$_{1-x}$\emph{Ln}$_{x}$FBiS$_2$ ($Ln =$La, Ce) with the tetragonal crystal structure has been synthesized and investigated\cite{LiSrF,LiCeF}. The parent compound SrFBiS$_2$ is a band insulator without detectable antiferrimagnetic transition or structure phase transition\cite{SrF,es}. Superconductivity with $T_c$ of $\sim$ 2.8 K has been induced by rare earth elements La or Ce doping into lattice, but its normal state exhibits the semiconducting-like behaviors even in the optimal superconducting sample\cite{CeFS,SSTLi}. The suppression of semiconducting behavior through increasing the carriers density was claimed to contribute the enhanced superconductivity\cite{maple2}. Therefore, most studies on the BiS$_{2}$-based system have mainly been focused on the effect of chemical substitution and superconducting transition temperature\cite{SCT}.

Pressure is considered as a clean method for adjusting the lattice parameter and the electronic band structure, which can enhance superconductivity and tune the normal state behaviors. Among the BiS$_2$-based superconductors, LaO$_{1-x}$F$_x$BiS$_2$ shows superconductivity below $T_c =$ 2.6 K at ambient pressure, but its $T_c$ reaches a maximum value of $\sim$ 10 K at $P =$ 2 GPa\cite{Maple}. The finding seems to be a universal feature, which is also observed in most BiS$_2$-based superconductors with a dramatic enhancement of $T_c$ under pressure\cite{Maple,Awana,EuF3244}. The reason may be hinted by a X-ray diffraction measurement which indicates that the system undergoes a structural phase transition from a tetragonal phase to a monoclinic phase at a critical pressure\cite{highXRD}. Hall effect measurements in Eu$_3$F$_4$Bi$_2$S$_4$ reveal also a change in electronic structure across the superconducting phase under pressure\cite{EuF3244}. Up to now, Sr$_{1-x}$\emph{Ln}$_{x}$FBiS$_2$ ($Ln =$ rare earth elements) has exhibited the unique properties but is less investigated\cite{LiSrF}. An interesting observation is that the La doping can induce a semiconductor-to-metal transition in the Sr$_{0.5}$La$_{0.5}$FBiS$_2$ system, accompanied with a sign changed Hall coefficient, even though its band structure is similar to that of LaO$_{1-x}$F$_x$BiS$_2$\cite{es}. In Sr$_{0.5}$Ce$_{0.5}$FBiS$_2$, the diluted-Ce ions order ferromagnetically at about 7.5 K, and coexist with superconductivity below 2.8 K\cite{LiCeF}. Therefore, the study of substitution effect for other magnetic rare earth elements such as Pr in SrFBiS$_2$ system is highly desirable, in particular for superconductivity, the normal state properties and its pressure effect.

In this paper, we report the successful synthesis of the Pr-doped Sr$_{1-x}$Pr$_{x}$FBiS$_{2}$ (0 $\leq x \leq$ 0.7) samples, and investigate their superconductivity, the detailed pressure effect and its normal state properties. We found that Pr-doping can immensely decrease the resistivity and induce superconductivity with the maximum $T_c$ of 2.7 K as x $\geq$ 0.4, but the normal state still remains the semiconducting-like behaviors.
$\chi(T)$ and $C(T)$ data suggest that those superconducting samples should exhibit the filamentary superconductivity because of the low shielding volume fractions and the absence of the superconduing transition jump in specific heat. According to our experimental results, the electronic state diagram can be obtained. An applied pressure can suddenly enchance the $T_c$, reaching 8.5 K at a critical pressure of $P_c =$ 1.5 GPa, and then increasing slightly to 9.7 K up to 2.5 GPa. Accompanied with superconductivity acrossing from low-$T_c$ phase to high-$T_c$ phase, the normal state resistivity undergoes a semiconductor-to-metal transition under pressure. Meanwhile, the thermal activation energy E$_g$ decreases gradually to zero at around $P_c$. The upper critical field ($B_{c2}(0)$) at high pressure is over ten times larger than that at ambient pressure. The ratio of $B_{c2}(0)/T_c$ dramatically increases from 0.6 T/K at $P =$ 0 to 2.6 T/K at $P =$ 2.5 GPa, in contrast to that observed in the isostructural EuFBiS$_2$. Those results imply that the nature of superconductivity both low-$T_c$ phase and high-$T_c$ phase maybe distinct in the present system,
the low-$T_c$ phase remains the filamentray superconductivity while the high-$T_c$ phase is the bulk superconductivity because of the structure phase transition.

\section{\label{sec:level1}Experiment}

The polycrystalline samples of Sr$_{1-x}$Pr$_{x}$FBiS$_{2}$ were synthesized by two-step solid state reaction method. The detailed synthesis methods can be found in the previous literature\cite{LiSrF}. Crystal structure characterization was performed by powder X-ray diffraction (XRD) at room temperature using a D/Max-rA diffractometer with CuK$_{\alpha}$ radiation and a graphite monochromator. Lattice parameters were obtained by Rietveld refinements. The (magneto)resistivity under several magnetic fields was performed with a standard four-terminal method covering temperature range from 0.5 to 300 K in a commercial Quantum Design PPMS-9 system with a $^{3}$He refrigeration insert. The temperature dependence of d.c. magnetization was measured on a Quantum Design SQUID-VSM-7T. Measurement of resistivity under pressure was performed up to 2.5 GPa on PPMS-9T by using HPC-33 Piston type pressure cell with the Quantum Design DC resistivity and AC transport options. The sample was immersed in a pressure transmitting medium (Daphne Oil) covered with a Teflon cap. Annealed Au wires were affixed to contact surfaces on each sample with silver epoxy in a standard four-wire configuration.

\section{\label{sec:level1}Results and Discussion}

\subsection{\label{sec:level1}Electronic phase diagram and superconductivity in Sr$_{1-x}$Pr$_{x}$FBiS$_{2}$ (0 $\leq x \leq$ 0.7)}

\begin{figure}
\includegraphics[width=14cm]{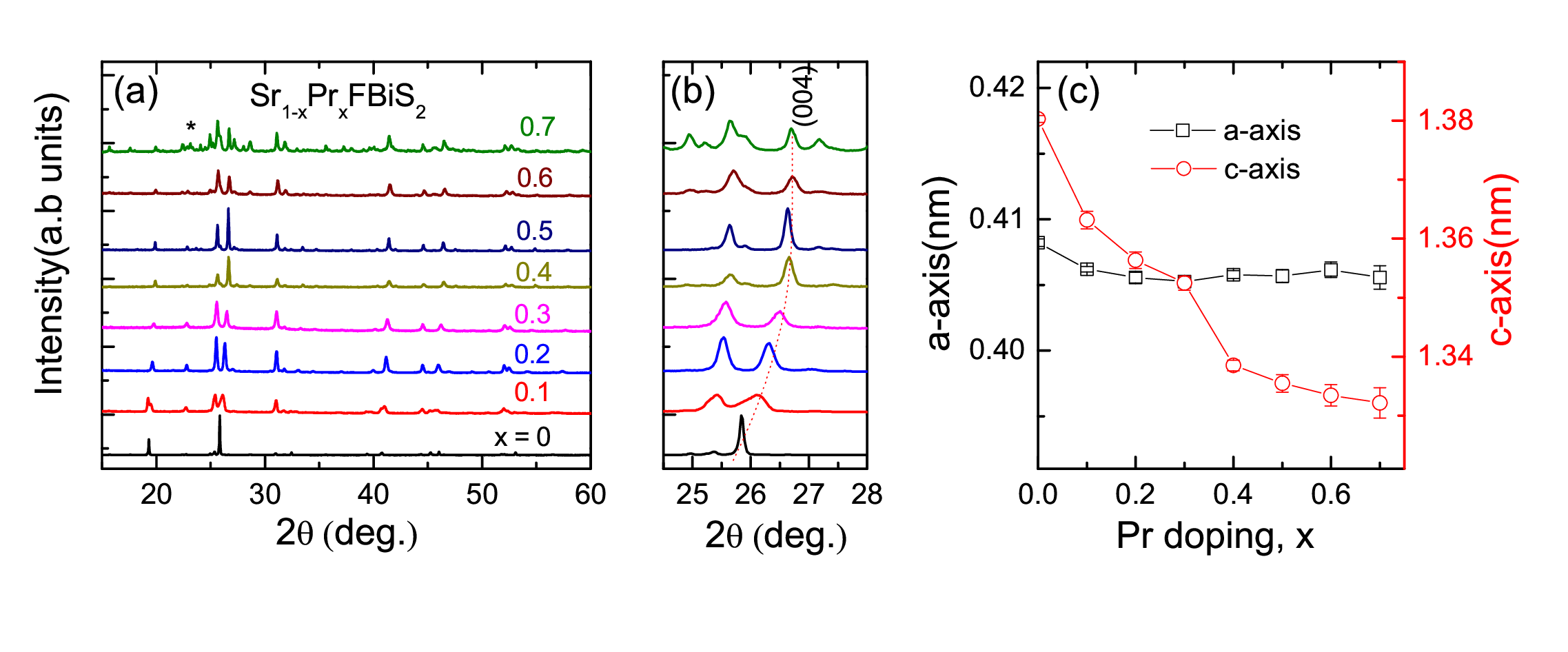}
\caption{\label{Fig.1} (color online). (a) Powder X-ray diffraction patterns for Sr$_{1-x}$Pr$_{x}$FBiS$_{2}$(0 $\leq x \leq$ 0.7) at room temperature. The $*$ peak positions designate the impurity phase of Bi$_{2}$S$_{3}$. (b) The enlarged picture of (004) peak; (c) Lattice constants vs. Pr doping contents.}
\end{figure}

Figure 1 shows the powder XRD patterns for the Sr$_{1-x}$Pr$_{x}$FBiS$_{2}$(0 $\leq x \leq$ 0.7) samples. Main diffraction peaks in these samples can be well indexed based on a tetragonal crystal structure with the P4/nmm space group. For x $\geq$ 0.5, extra tiny peaks arising from impurity phase of Bi$_{2}$S$_{3}$ can be observed\cite{BiS}, and its content gradually increases with the Pr doping. About 70\% Pr doping produces the much more impurities phase, reaching the limit of the solid solubility of Sr/Pr. An enlarged picture for (004) peak is shown in figure 1b. This peak obviously shifts to the high angles but remains its position as x is over 0.6, implying that the Pr succeeds in doping into the Lattice and its maximum doping concentration closes to 0.7. Their lattice constants refined by the Rietveld structural analysis methods are plotted in figure 1c. Compared with those of the mother compound SrFBiS$_{2}$\cite{LiSrF}, the a-axis decreases very slowly and almost unchanges in the high doping regime, while the c-axis rapidly decreases with the Pr doping concentrations. As a result, the cell volume has a significant decrease. The remarkable shrink in lattice parameters indicates the successful substitution of Sr by Pr, similar to the case of Ce-doped in Sr$_{0.5}$Ce$_{0.5}$FBiS$_{2}$ system\cite{LiCeF} and LaO$_{1-x}$F$_{x}$BiS$_2$ system\cite{LaFS}.

\begin{figure}
\includegraphics[width=14cm]{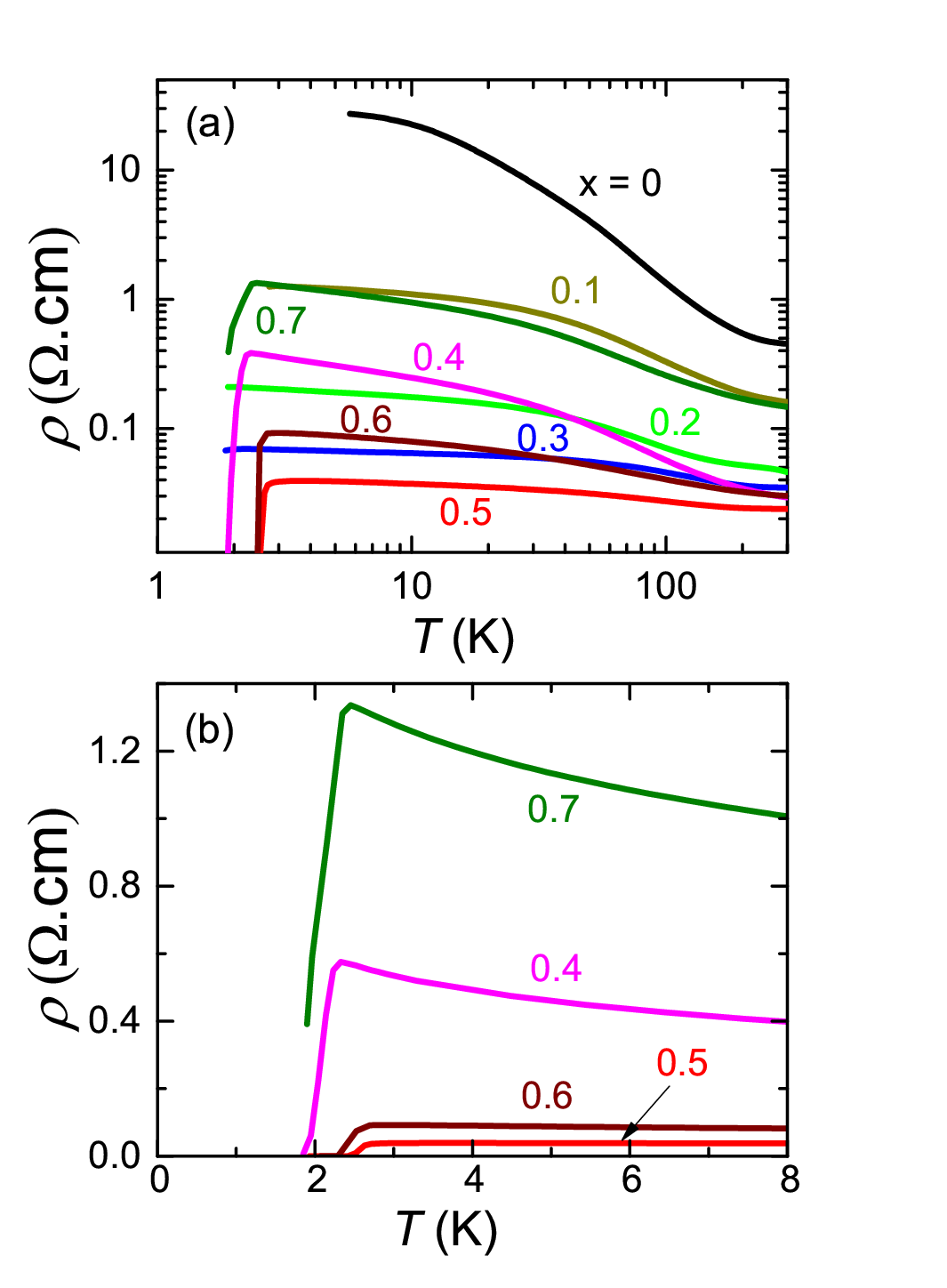}
\caption{\label{Fig.2} (color online). (a)Temperature dependence of resistivity ($\rho$) for the Sr$_{1-x}$Pr$_{x}$FBiS$_{2}$ (0 $\leq x \leq$ 0.7) samples. (b) An enlarged plot of resistivity around $T_{c}$.}
\end{figure}

Temperature dependence of resistivity $\rho$(T) in Sr$_{1-x}$Pr$_{x}$FBiS$_{2}$ is mapped in figure 2a. For the parent compound, resistivity exhibits a semiconducting-like behavior down to low temperatures. The resistivity value in the whole temperature regime decreases sharply with Pr doping to 0.3 becasue of the introduction of extra electrons. As $x \geq$ 0.4, a sharp superconducting transition is observed at low temperatures. The maximum $T_c$ of 2.7 K is observed at $x =$ 0.5 in figure 2b, whose normal state resistivity exhibits the minimum values but remains the semiconducting-like character. Such feature seems to be universal in the BiS$_2$-based superconductors\cite{LaFS,CeFS}. Increasing Pr doping contents to 0.7, the normal state resistivity reversely increases, and $T_c$ has a significant decrease without zero resistivity down to 2 K. By fitting with the thermal activation formula $\rho(T)=\rho_0 \exp(E_a/k_B T)$ for the temperature range from 120 K to 300 K, the thermal activation energy ($E_g$), as shown in figure 5, decreases gradually with Pr doping (The detail will be discussed below).

\begin{figure}
\includegraphics[width=14cm]{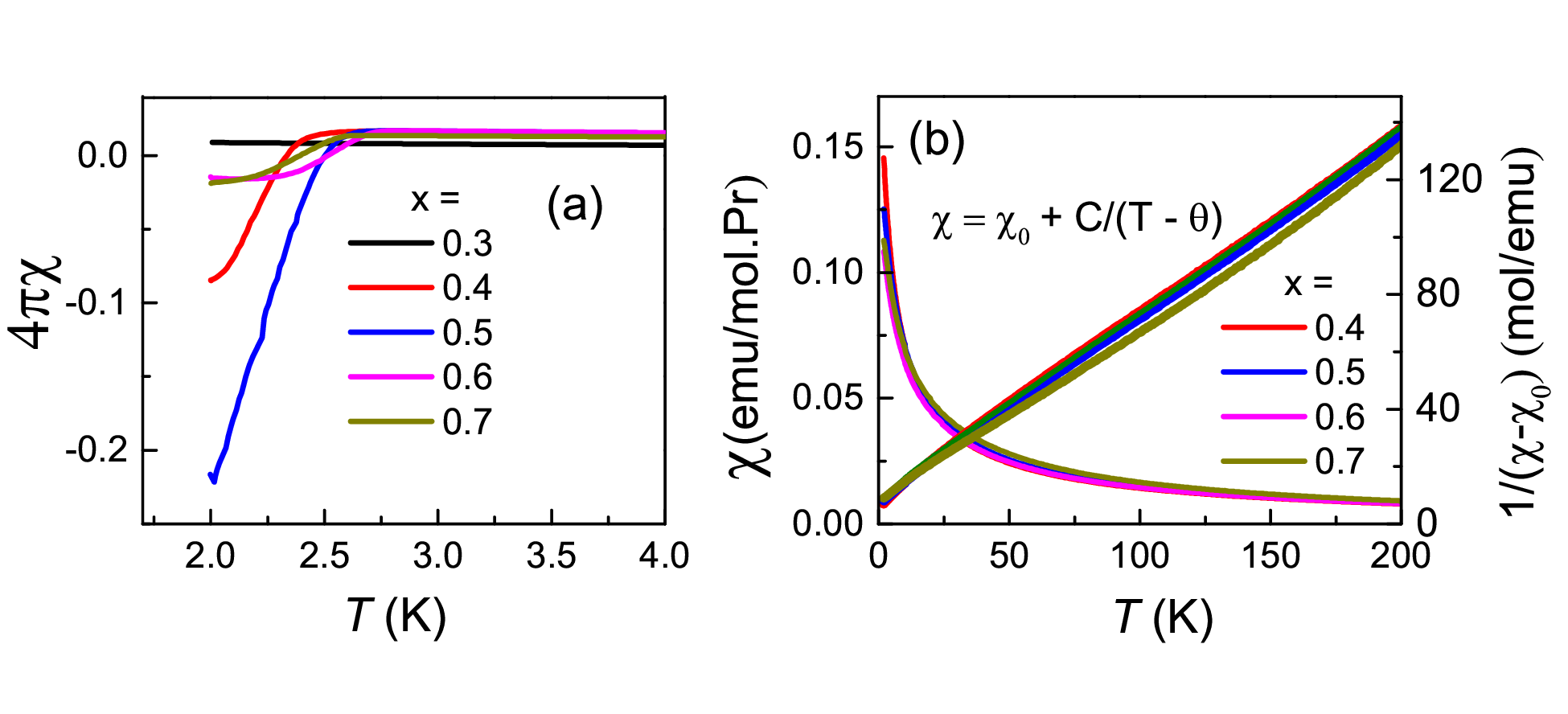}
\caption{\label{Fig.2} (color online). Temperature dependent magnetic susceptibility under (a) 5 Oe and (b) 1000 Oe magnetic fields with ZFC mode for Sr$_{1-x}$Pr$_{x}$FBiS$_{2}$ samples}
\end{figure}

\begin{table}
\begin{center}
\caption{\label{Tab1} Superconducting transition tenperature $T_c$ derived from $\rho(T)$, electrnoic coefficient $\gamma_0$, $\Theta$ and magnetic entropy $S_m$ obtained from the zero field specific heat data for Sr$_{1-x}$Pr$_{x}$FBiS$_{2}$ samples. As a comparision, the corresponding parameters for Sr$_{0.5}$(La,Ce)$_{0.5}$FBiS$_2$\cite{LiSrF,LiCeF} and EuFBiS$_2$\cite{EuFBiS} samples are also summarized in table 1. \\}

\begin{tabular}{cccccccccc}
\hline
$x$                              &   0.4  &  0.5   &   0.6   &   0.7 & Sr$_{0.5}$La$_{0.5}$FBiS$_2$ & Sr$_{0.5}$Ce$_{0.5}$FBiS$_2$& EuFBiS$_2$& \\ \hline
$T_c$ (K)                        &  2.23  &  2.72  &   2.65  &  2.44 &   2.8  &   2.7  &  0.3     \\
$\gamma_0$ [mJ/(mol$\cdot$K$^2$)]&  104   &  154   &   128   &  162  &   1.42 &  117.2 &  73.3    \\
$\Theta_D$(K)                      &  206   &  215   &   196   &  207  &   265  &  220   &  201   \\
$\mu_{eff}$($\mu_B$)             &  3.45  &  3.47  &   3.49  &  3.52 &   -    &  2.53  &  7.2     \\
$S_{mag}$   [J/(mol$\cdot$K)]    &  0.76  &  0.45  &   0.42  &  0.4  &   -    &  2.7   &  12.4       \\
\hline
\end{tabular}
\end{center}
\end{table}

Note that in the previous literature\cite{Awana2} superconductivity was not seen in the Sr$_{0.5}$Pr$_{0.5}$FBiS$_{2}$ sample at ambient pressure. To determine the bulk superconductivity in our sample, the d.c. magnetic susceptibility with zero field cooling (ZFC) modes under 5 Oe is shown in figure 3a. Below $T_c$, the diamagnetic signal can be clearly observed as $x \geq$ 0.4. From ZFC data, the estimated superconducting shielding volume fractions for all superconducting samples are below 25\%, implying that those samples may exhibit filamentary/weak superconductivity through Pr doping, similar to the case of the LaO$_{0.5}$F$_{0.5}$FBiS$_{2}$\cite{LaFS}. The relatively small volume fraction is commonly found in the BiS$_2$-based polycrystalline samples, which is ascribed to a low-$T_c$ phase\cite{LaFS,Lowsignal} or in-plane disorder\cite{Inplane}.

Temperature dependence of magnetic susceptibility for $0.4 \leq x \leq 0.7$ samples is shown in figure 3b. Overall, the $\chi(T)$ of those samples rapidly increases and shows the Curie-Weiss behaviors with cooling temperature. No anomaly associated with the magnetic ordering of Pr moments down to 2 K is observed. In contrast, Sr$_{0.5}$Ce$_{0.5}$FBiS$_{2}$ displays the ferromagnetic ordering at 7.5 K\cite{LiCeF} and EuFBiS$_{2}$ exhibits the antiferromagnetic transition at 2.3 K\cite{EuFBiS}. Such divergency implys that $Pr^{3+}$ ions with an interger anglular momentum (j =4) may result a Kramers non-magetic ground state in those samples, similar to the case of PrNiAsO superconductor\cite{PrNiAsO}. According to the Curie-Weiss law: $\chi(T)=\chi(0)+ \frac{C}{T-\theta}$, where $\chi_{0}$ denotes the temperature-independent term, $\emph{C}$ is the Curie-Weiss constant and $\theta$ denotes the paramagnetic Curie temperature, we therefore fit $\chi(T)$ above 50 K. By subtracting the temperature-independent term ($\chi_{0}$), the $(\chi -\chi_{0})^{-1}$ shows a linear T-dependence in figure 3b, suggesting a reliable fitting result. The fitted effective moments $\mu_{eff}(Pr)$ for $x = 0.4-0.7$ samples, as shown in table 1, are closed to the theoretical value of 3.57 for a free Pr$^{3+}$ ion.


\begin{figure}
\includegraphics[width=14cm]{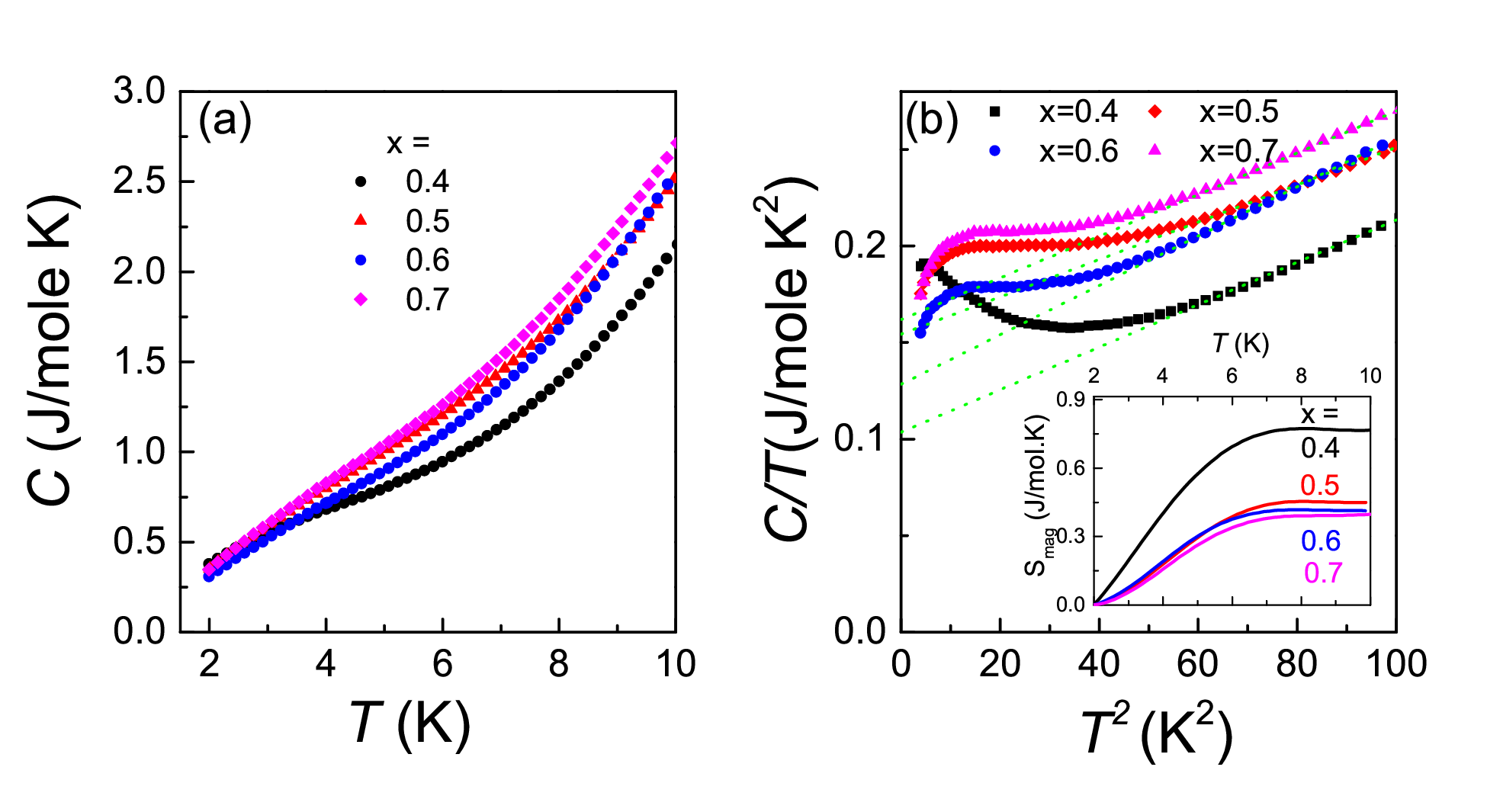}
\caption{\label{Fig.2} (color online). (a) Low temperature specific heats for Sr$_{1-x}$Pr$_{x}$FBiS$_{2}$ ($x= 0.4-0.7$) samples. (b) the $C/T$ vs. $T^2$ in the main plane; the inset shows the magnetic entroy for those samples.}
\end{figure}

Specific heat data of $x = 0.4 - 0.7$ superconducting samples below 10 K are plotted in figure 4a. No peak origining from superconducting transitinon or magnetic ordering of Pr moments can be observed in the temperature regime, confirming that Pr doping may induce the filamentary superconductivity but does not form the long range magnetic ordering. The total specific heat can be written as: $C(T)=\gamma_0 T + \beta T^3 + C_{mag}$, where $\gamma_0$ and $\beta$ are coefficients of the electron and phonon contributions, while $C_{mag}$ is the Pr-$4f$ magnetic term. Considering the negligible magnetic contributions of Pr-$4f$, we fit the data and plot the $C(T)/T$ vs. $T^2$ in figure 4b. Above $\sim$ 7 K, the good linear fitting for all samples is observed and yields the Sommerfeld coefficient and Debye temperature ($\Theta_D$) listed in table 1. $\Theta_D$ value in the Pr-doping samples is much smaller than those of Sr$_{0.5}$(La,Ce)$_{0.5}$FBiS$_2$ (265 K, 220K), consistent with the substitution of lighter Sr ions by heavier Pr ions. The derived $\gamma_0 \sim $ 154 mJ/(mol$\cdot$K$^2$) for $x =$ 0.5 is surprisingly large, 108 times of that of Sr$_{0.5}$La$_{0.5}$FBiS$_2$ (1.4 mJ/(mol$\cdot$K$^2$)), and over twice times of that of Sr$_{0.5}$Ce$_{0.5}$FBiS$_2$ (58.6 mJ/(mol$\cdot$K$^2$)) and EuFBiS$_2$ (73.3 mJ/(mol$\cdot$K$^2$)). Note that the similar large $\gamma_0$ has been reported in the iso-structure PrO$_{0.5}$F$_{0.5}$BiS$_2$ superconductor (286.36 mJ/(mol$\cdot$K$^2$))\cite{PrOFBiS}. The great enhanced Sommerfeld coefficient suggests the mainly contributions originating from the hybridization between conduction electrons and the Pr-$4f$ electrons. The magnetic entropy $S_{mag}$ can be obtained by the integration of $C_{mag}/T$ over T up to 10 K, as shown in the inset of figure 4b. The released entropy is just equal to about 2.2-4.2\% of the expected value for $J = 4$ in the present samples. Unlike the FM order of Ce$^{3+}$ ions in Sr$_{0.5}$Ce$_{0.5}$FBiS$_2$ and the AFM of Eu$^{2+}$ in EuFBiS$_2$ with the large magnetic entropy, the Pr-doped samples exhibits the low magnetic entropy ($S_{mag}$) in table 1, reasonablely consistent with the non-magnetic ground state in the present samples.

\begin{figure}
\includegraphics[width=14cm]{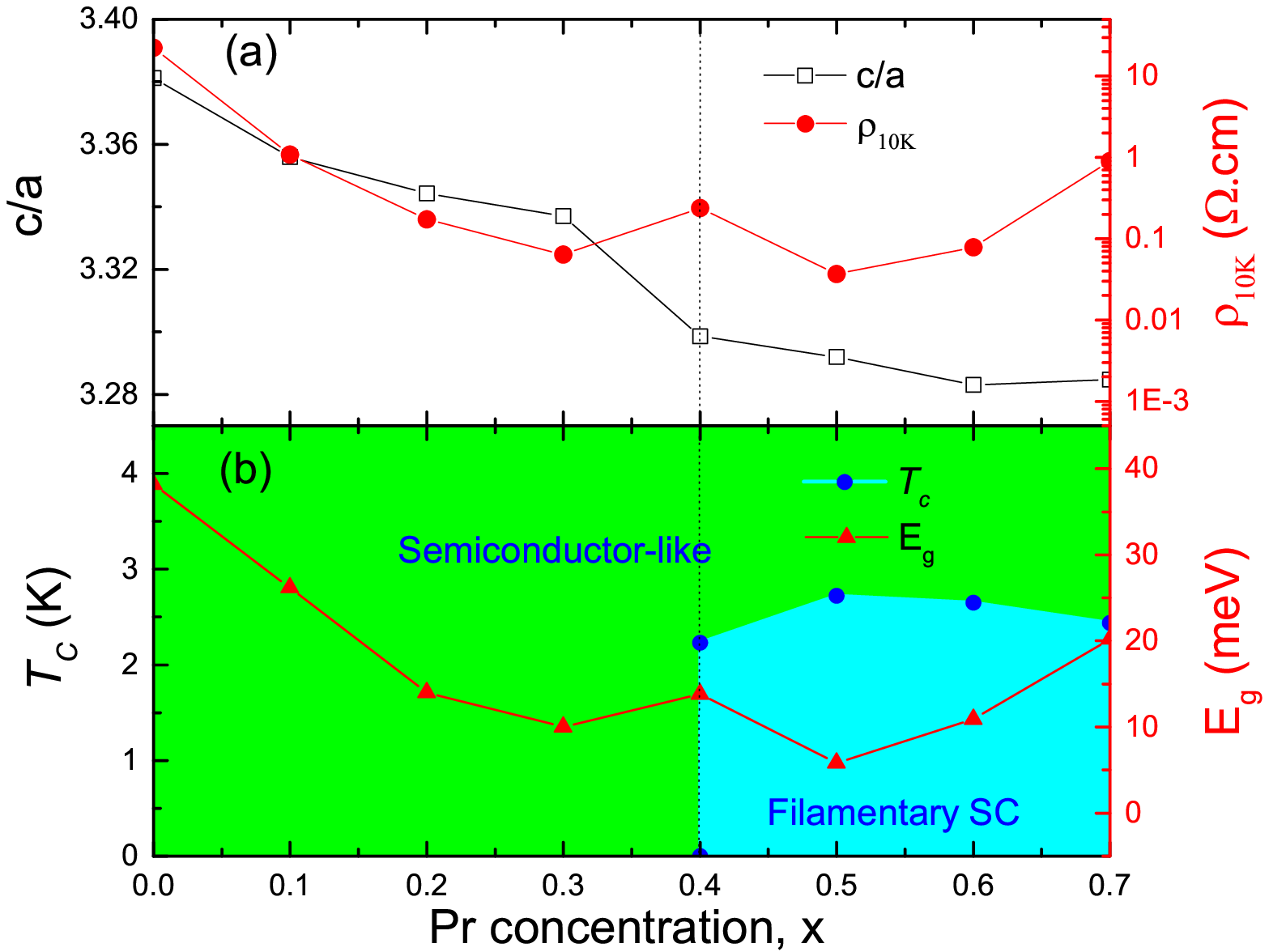}
\caption{\label{Fig.2} (color online). Phase diagram of Sr$_{1-x}$Pr$_{x}$FBiS$_{2}$ system with Pr-doing concentrations.}
\end{figure}

The electronic phase diagram for Pr-doped system can thus be summarized in figure 5. In the low Pr-doping concentrations (x $\leq$ 0.3), the sample shows the semiconductor-like behaviors in the whole temperature regime. Superconductivity starts to emerge as only $x$ is equal to 0.4. For x $\geq$ 0.4, the normal state exhibits the semiconducting-like character, and the $T_c$ exhibits the maximum value at x $=$ 0.5 and then decreases slightly with the x increasing to 0.7. On the other hand, the thermal activation energy E$_g$ estimated from the resisitivity data decreases rapidly in the low doping regimes ( $\leq$ 0.3), and shows an abnormal increase at x = 0.4. The minimum E$_g$ is observed for x $=$ 0.5 whose T$_c$ is the highest. Further increasing x to 0.7, the E$_g$ reversely increases again. A combination of the change of $T_c$ and E$_g$ with the Pr-doping concentrations suggests that superconductivity competes with the semiconducting normal state in the present system and 40\% Pr-doping contents may induce the Fermi surface change or a possible Lifshize transition, leading to the appearance of superconductivity. Such change is also observed in the Sr$_{1-x}$La$_{x}$FBiS$_{2}$ system\cite{SSTLi}. Correspondingly, the ratio of lattice parameters $\frac{c}{a}$ shows a slight drop and the resisitivity $\rho_{10K}$ exhibits the anomaly at $x =$ 0.4 in figure 5a, implying that a possible slight change of the lattice may partly contribute to the emergence of superconductivity in this system. As a comparison, in Sr$_{1-x}$La$_{x}$FBiS$_{2}$ system\cite{SSTLi}, La-doping can result in an rapid increase of the metallic conductivity and induce the bulk superconductivity with La-doping, but in Sr$_{1-x}$Pr$_{x}$FBiS$_{2}$ the semiconducting-like normal state is rather robust even $x =$ 0.7 and the evidences of bulk superconductivity seems to be rather absent. Such feature
can be found in some BiS$_2$-based superconductors\cite{LaFS,NdFS} and the reason can be ascribed to the weak overlap of Bi-6p and S-p orbits associated with the robust semiconducting-like normal state in resistivity\cite{Inplane,Inplane2}.

\subsection{\label{sec:level1}Effect of External Pressure on Superconductivity of Sr$_{0.5}$Pr$_{0.5}$FBiS$_{2}$}

\begin{figure*}
\includegraphics[width=8cm]{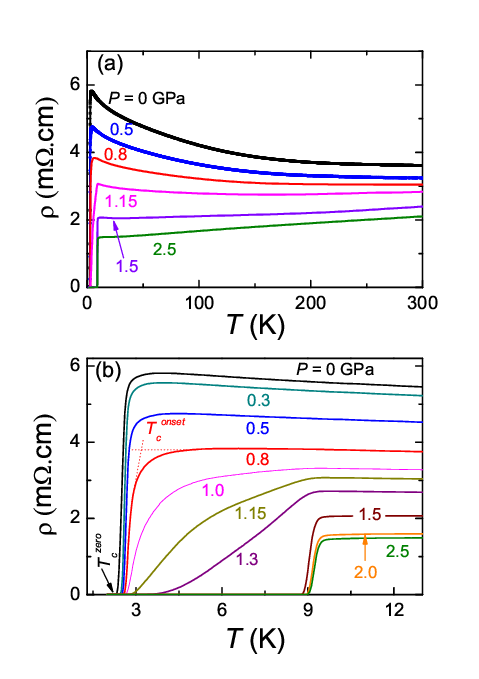}
\includegraphics[width=8cm]{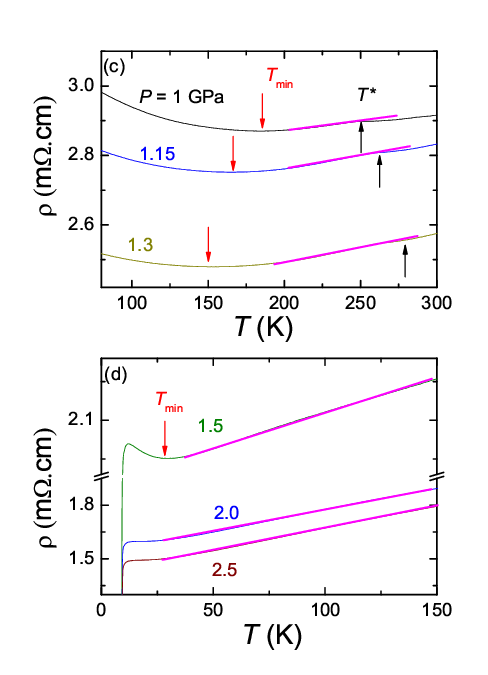}
\caption{\label{Fig.4}(color online) (a) Temperature dependence of resistivity under several representative pressures for Sr$_{0.5}$Pr$_{0.5}$FBiS$_{2}$ sample from 2 K to 300 K. (b) An enlarged view of superconducting transition in resistivity below 13 K. (c) and (d) enlarged plots of $\rho(T)$ in the high temperature regime}
\end{figure*}

A strategy to enhance the overlap of orbital is high pressure effect. Several cases in the BiS$_2$ systems\cite{Maple,Awana,EuF3244} have been achieved to enhance superconductivity under pressure. We thus perform the pressure measurements on the optimal doping ($x =$ 0.5) smaple. Temperature dependence of resistivity at various pressures is summarized in figure 6. Fig. 6a shows $\rho (T)$ upon increasing pressure to 2.5 Gpa. The semiconducting behavior is clearly observed at lower pressures, gradually suppressed with increasing pressure and finally turns to metallic at 2.5 Gpa. Accompanied with a decrease of resistivity in the normal state, the $T_c$ is enhanced under pressure. An expanded view of resistivity around $T_c$ is shown in figure 6b. For $P \leq$ 0.8 GPa, $T_c^{onset}$, as illustrated in figure 6 (b), has a slow increase with increasing pressures. As pressure is in the range 0.8 to 1.3 GPa, the superconducting transition broadens significantly, and two successive superconducting transitions for 1.15 GPa are detected at 5.5 K and 8.5 K, respectively. Further increasing pressure to 1.5 GPa, the transition becomes sharp again and $T_c^{onset}$ is significantly enhanced to about 9 K. For higher pressure above 1.5 GPa, $T_c^{onset}$ goes up slowly with pressure, and reaches 9.7 K at 2.5 GPa. The similar pressure effect was reported in the iso-structure EuFBiS$_2$ and LaO$_{1-x}$F$_{x}$BiSe$_2$ superconductors\cite{YuanHQ,WHHSe}, whose metallic conductivity can be enhanced under pressure.

In contrast, a semiconductor-to-metal transition can be observed with increasing pressure in the present Sr$_{0.5}$Pr$_{0.5}$FBiS$_{2}$ sample. As shown in figure 6c and d, at 1.0 GPa the resistivity exhibits a metallic behavior at high temperature, and then undergoes a metal to semiconductor transition around a characteristic temperature $T_{min}$. With further increasing pressure, $T_{min}$ gradually shifts to lower temperature and finally vanishes above 1.5 GPa. The similar feature was reported in Sr$_{1-x}$La$_{x}$FBiS$_2$ system, where the semiconductor-metal transition is induced by the La doping\cite{SSTLi,JPSJ}. The first-principle calculation\cite{CeCalculation} has proposed that \emph{Ln}O$_{1-x}$F$_{x}$BiS$_2$ will undergo a semiconductor-metal transition under pressure. On the other hand, accompanied with the enhancement of metallicity under pressure, the resistivity obviously shows a linear temperature-dependence above $T_{min}$, which is different to the case of EuFBiS$_2$ under pressure\cite{YuanHQ}. Note that as pressure is in the range from 1.0 to 1.3 GPa, a kink in resistivity about 250 K, as illustrated in figure 6c, is clear observed. A similar feature was observed in the EuFBiS$_2$ compound and was tentatively ascribed to the possible charge-density wave transition\cite{EuFBiS}.

\begin{figure}
\includegraphics[width=16cm]{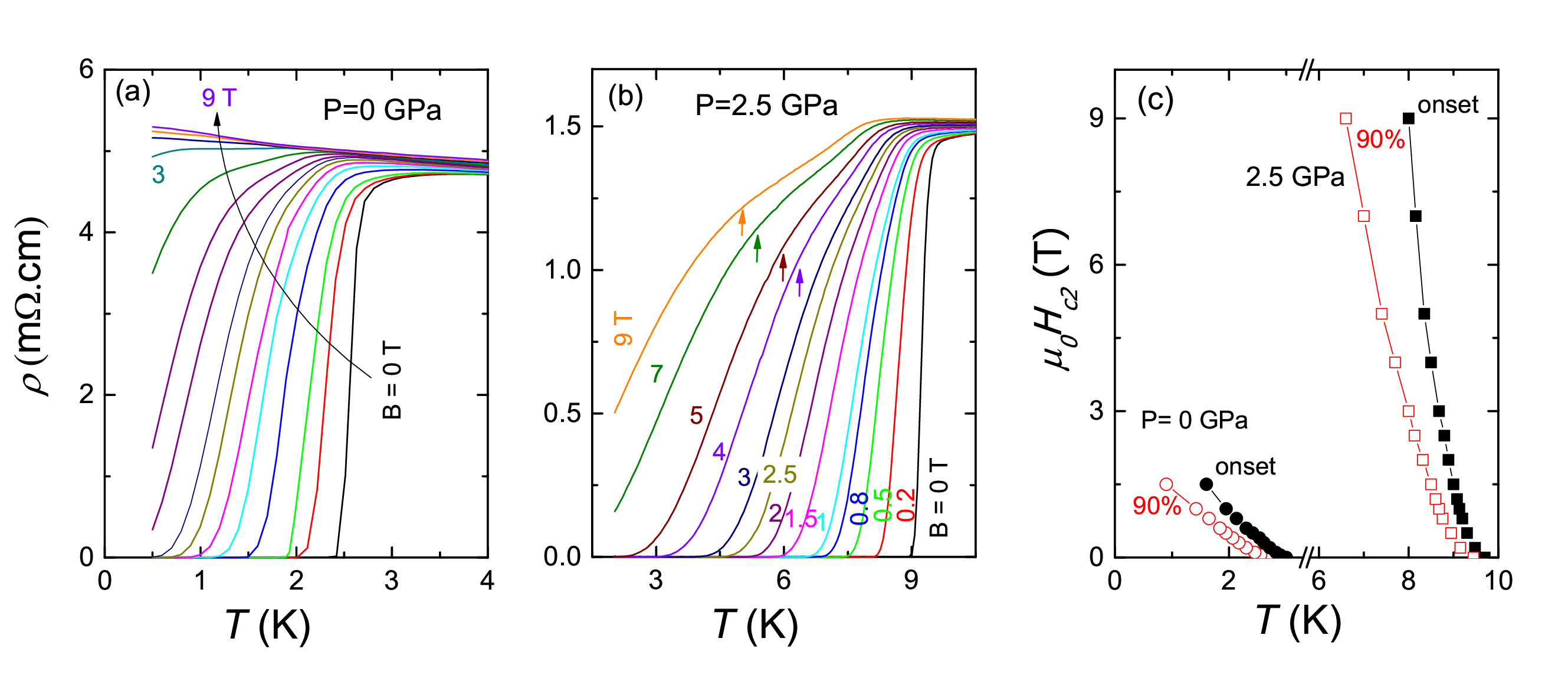}
\caption{\label{Fig.4}(color online) (a) Superconducting transition vs. T under various magnetic fields at ambient pressure and (b) 2.5 GPa. (c) Upper critical field  $\mu_0$$H_{c2}(T)$-T phase diagram at ambient pressure and 2.5 GPa, respectively. The $T_c^{onset}$ and $T_c^{90\%}$ are determined from the onset superconducting transition and 90\% of the normal state value in $\rho(T)$, respectively. }
\end{figure}

Temperature dependence of resistivity under various magnetic fields at ambient pressure and 2.5 GPa are plotted in figure 7. At ambient pressure, the sharp superconducting transition slowly brodens and the $T_{c}$ gradually shifts to below 2 K as the external magnetic fields are increased from zero to 3 T in figure 7a. For $B \geq$ 5 T, superconductivity completely disappears and resistivity recovers a weak semiconducting-like feature, similar to the case of Bi$_{4}$O$_{4}$S$_{3}$ system\cite{WenHH}. In the case of P = 2.5 GPa, the high $T_c$ of 9.7 K is observed at zero field. As the magnetic field is applied, $T_c$ slightly shifts to lower temperatures and the transition gradually broadens. As magnetic fields exceed 4 T, a shoulder in $\rho(T)$ curve is detected, as marked using arrow symbol in figure 7b. A similar character has been reported in the high pressure phase of EuFBiS$_2$ and Eu$_3$F$_4$Bi$_2$S$_4$\cite{YuanHQ,EuF3244}, which was interpreted as the anisotropy in the upper critical field. However, superconductivity seems to be rather robust keeping strong signals above 2 K even though the field is up to 9 T. As a comparison, in both EuFBiS$_2$ and Eu$_3$F$_4$Bi$_2$S$_4$ at high pressure\cite{YuanHQ,EuF3244}, a magnetic field of about 3.5 T can completely destroy superconductivity. The upper critical fields of P = 0 GPa and 2.5 GPa are displayed in figure 7c, where the criterions of $T_c^{onset}$ and $T_c^{90\%}$ are employed to determined the $\mu_0H_{c2}$. The zero temperature limit estimated by the WHH formula ${H_{c2}(T) = -0.69T_c|{\frac{\partial{H_{c2}}}{\partial{T}}|_{T_c}}}$ is about 1.6 T and 25 T for ambient pressure and 2.5 GPa, respectively. As a result, the value of $B_{c2}(0)/T_c$ dramatically increases from 0.6 T/K at $P =$ 0 to 2.6 T/K at $P =$ 2.5 GPa, implying possible different origin of superconductivity in the corresponding superconducing phase. In contrast, the iso-structural EuFBiS$_2$ has a comparable $T_c$ of 9 K under 2.4 GPa, but its upper critical field is rather small, and $B_{c2}(0)/T_c$ value abnormally decreases under pressure\cite{YuanHQ}.

According to the above experimental results, the phase diagram of Sr$_{0.5}$Pr$_{0.5}$FBiS$_{2}$ under pressure is summarized in figure 8. For lower pressure, $T_c$ has a slight increase with increasing pressure. As pressure crosses a critical value of $P_c \simeq$ 1.5 GPa, $T_c$ dramatically increases from 3 K to around 9 K, and a high-$T_c$ SC phase emerges at higher pressure. Two distinct superconducting regions including the low-$T_c$ phase(SC1) below $P_c$ and the high-$T_c$ phase(SC2) above $P_c$ are clearly distinguished, sharing a common feature in the most BiS$_2$-based superconductors\cite{Maple,EuF3244,YuanHQ}. Such feature is likely related to the structure transition from a tetragonal phase to a monoclinic phase at a critical pressure ($P_c$)\cite{highXRD}. By coincidence, both the $E_g$ and $T_{min}$ decrease suddenly and tend to approach zero across the critical pressure $P_c$. We thus suggest that low-$T_c$ phase may be still ascribed to the filamentary superconductivity where the tetragonal phase dominates below $P_c$, but the higher pressures ($ > P_c$) maybe induce the bulk high-$T_c$ phase because of the occurence of the structure phase transition.

\begin{figure}
\centering
\includegraphics[width=14cm]{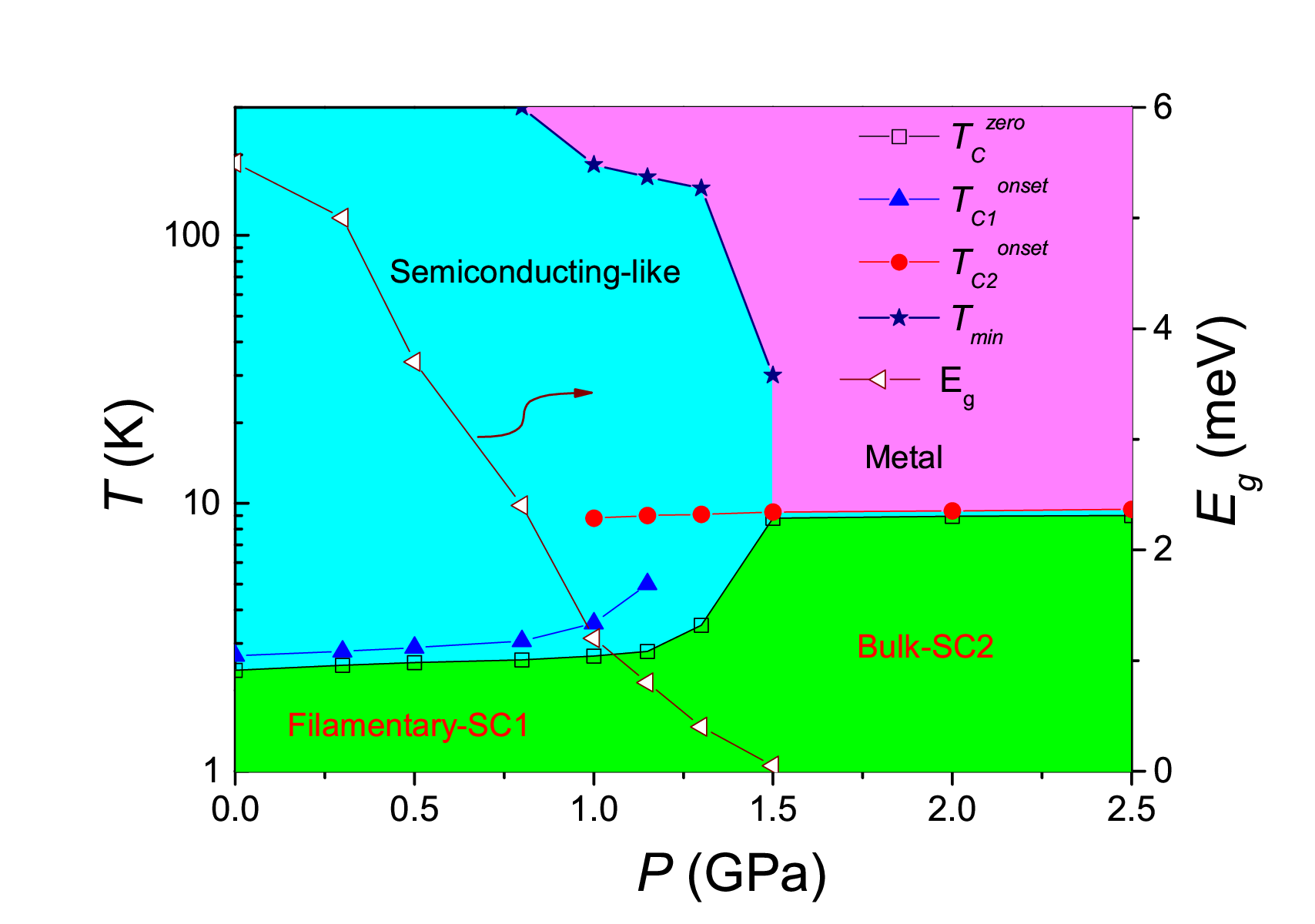}
\caption{\label{Fig.8}(color online)  Phase diagram of pressure vs. Temperature for Sr$_{0.5}$Pr$_{0.5}$FBiS$_{2}$ sample. $T_c^{onset}$ and $T_c^{zero}$ represent the onset superconducting transition and zero resistivity in $\rho(T)$, respectively. $T_{min}$ characterizes the semiconductor-metal transition from resistivity measurement. The thermal activation energy $E_g$ is fitted for the temperature range from $T_c$ to $T_{min}$ by the mentioned formula in text.}
\end{figure}

\section{\label{sec:level1}Conclusion}

In summary, we have reported the BiS$_{2}$-based Sr$_{1-x}$Pr$_{x}$FBiS$_{2}$ system by the measurements of resistivity, magnetic susceptibility, specific heat and presure effect. Our experimental results show that Pr doping can significantly decrease the sample's resistivity and induce superconductivity with the maximum $T_c$ of 2.7 K at $x =$ 0.5, but its normal state resistivity remains the semiconducting-like behaviors.
However, both the low superconducting shielding volum fractions in $\chi(T)$ and the absence of the superconducting transition jump in $C(T)$ obviously point to a filamentary superconductivity in the Pr-doping samples. Applying a pressure can abruptly enhance $T_c$ from 2.7 K at ambient pressure to about 8.5 K at a critical pressure $P_c =$ 1.5 GPa, reaching about 9.7 K at 2.5 GPa. Meanwhile, the normal state resistivity undergoes a semiconductor to metal transition, and both $T_{min}$ and $E_g$ decrease and almost approach zero at $P_c $. Two superconducting states including the low-$T_c$ phase and the high-$T_c$ phase are separated by a coexisting area at the near critical pressure $P_c$. We believe that the low-$T_c$ phase is from the filamentary superconductivity because of weak orbitals overlap while the bulk superconductivity may be induced in the high-$T_c$ phase due to the structure transition from tetragonal phase to trigonal phase under pressure. Moreover, the upper critical field ($\mu_0H_{c2}(0)$) estimated in the high-$T_c$ phase at 2.5 GPa is over ten times larger than that of at ambient pressure. As a result, the ratio of $B_{c2}(0)/T_c$ dramatically increases from 0.6 T/K at $P =$ 0 to 2.6 T/K at $P = $2.5 GPa. All these results suggest that the two different superconducting phases may possess the different origin and deserve to be investigated by further experiments.

\section*{Acknowledgments}
This research was supported in part by the NSF of China (under Grants No. 11474082, 61401136 and 61376094), Natural Science Foundation of Zhejiang Province (LY18F010019), and QianJiang talents program of Zhejiang Province. Yu-Ke Li was supported by an open program from Wuhan National High Magnetic Field Center (2016KF03). Jie Cheng was supported by the General Program of Natural Science Foundation of Jiangsu Province of China (No. BK20171440). Y. Luo acknowledges the support from the 1000 Youth Talents Plan of China..

\end{document}